\documentclass[12pt]{iopart}
\usepackage{graphicx}
\usepackage{amssymb}
\usepackage{xcolor}

\begin{document}

\title{SQUID current sensors with an integrated thermally actuated input current limiter}

\author{R K\"orber$^1$, P Krzysteczko$^1$, M Klemm$^2$, T Liu$^1$, J-H Storm$^3$ and J Beyer$^1$}

\address{$^1$ Physikalisch-Technische Bundesanstalt (PTB), Abbestrasse 2-12, D-10587 Berlin}
\address{$^2$ Magnicon, Division Berlin, Abbestrasse 2-12, D-10587 Berlin}
\address{$^3$ Empit, Alt-Biesdorf 11, D-12683 Berlin}

\ead{Rainer.Koerber@ptb.de}

\vspace{10pt}
\begin{indented}
\item[\today]
\end{indented}

\begin{abstract}
The development of SQUID current sensors with a thermally actuated input current limiter (TCL) integrated into the input circuit of the sensor is presented. The TCL is based on an unshunted Josephson Junction (JJ) series array, around which meanders a galvanically isolated, but tightly, thermally coupled, resistive heater element. By applying a current to the heater, the JJ critical currents can be reduced or completely suppressed, while the other on-chip SQUID circuit elements remain unaffected. The functional parameters of the TCL are determined by direct transport measurements and by static and dynamic flux coupling tests via the input circuit. In liquid Helium, a heater power, which reliably suppresses the critical current of the JJ array to zero, of $\sim$1.5~mW was needed, with the TCL normal state resistance being at 500~$\Omega$. In this configuration, typical TCL switching times below 10~$\mu$s were observed in the direct transport measurements.
	
The TCL can be used to temporarily or even permanently disable malfunctioning channels in multichannel SQUID magnetometer systems, when feedback into their otherwise superconducting input circuits is not possible. In doing so, significant signal distortions in neighbouring channels from screening currents in these input circuits are avoided. Furthermore, by completely suppressing and restoring the TCL critical current, dc offset currents in the superconducting input circuits can be prevented. This is relevant, for instance, in SQUID-based spin precession experiments on hyperpolarized noble gases, in which even dc currents of a few $\mu$A in the input circuit can lead to significant magnetic field distortions in the adjacent sample region contributing to the transverse spin-spin relaxation rate $1/T_{2}$.

\end{abstract}

\vspace{2pc}
\submitto{\SUST}
\maketitle
\ioptwocol	

\section{Introduction}
The superconducting quantum interference device (SQUID) forms the basis of many highly sensitive measurement devices~\cite{Clarke2006} and one of the most common arrangements is the current sensing SQUID. Here, the SQUID is usually inductively coupled to an on-chip input coil which, if used as a magnetometer, can be connected to a superconducting pick-up coil. This broadband detection scheme is deployed in biomagnetism, where multichannel systems map the magnetic fields emanating from the human body. Other applications require the use of pulsed fields prior to a detection period, e.g., ultra-low field magnetic resonance imaging (ULF MRI)~\cite{Kraus2014} or magnetorelaxometry (MRX)~\cite{Arsalani2022}. In these cases, the SQUID circuitry must be protected from potentially large currents in the input circuit which tend to trap flux in the form of vortices in the superconducting on-chip structures. Arbitrary motion, such as thermally activated agitation of such vortices, would lead to excess low frequency noise which degrades the performance of the device.

One method to mitigate this problem is to limit the current in the superconducting input circuit, by incorporating either a series Josephson Junction (JJ) array~\cite{Hilbert1985a} or a series SQUID array~\cite{Drung2007}. In both cases, the signal current in the superconducting input circuit cannot exceed the critical current $I_{\rm{c}}$ of the JJ- or SQUID series array. Even though $I_{\rm{c}}$ can also be modulated externally by applying a magnetic field to this type of current limiter, it cannot be set precisely to zero. Another approach is to thermally drive part of the superconducting input circuit leads into its normal state. This scheme is used, for instance, in a so-called cryoswitch~\cite{Zotev2007} (Supracon SW1). The cryoswitch, when not actuated, does not restrict the input current, though it could be combined with a JJ- or SQUID-based input current limiter. 

In this work, those approaches are combined in a design that employs a series array of JJs as well as a resistive heater. As a result, a compact, fully integrated, thermally actuated current limiter (TCL) is presented which offers a well-defined low critical current and the ability to set $I_{\rm{c}}={}$0.

\section{Design of thermally activated switch}

\begin{figure}[!t]
\centering
\includegraphics[width=0.94\columnwidth]{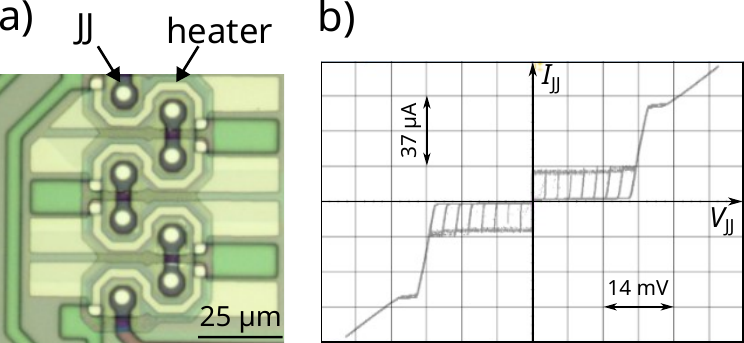}
\caption{a) Picture of the TCL showing the 8 JJs together with the galvanically isolated heater (taken before the final deposition of the top Nb electrode). b) Example of a measured $I$-$V$ curve of the series array of 8 JJs.}
\label{fig:fig_1}
\end{figure}

The TCL consists of a series array of 8 unshunted JJs, around which meanders a galvanically isolated heater, formed by a thin-film resistor as shown in figure~\ref{fig:fig_1}a). The window-type JJs are fabricated using the established Nb-Al(AlO$_{\rm{x}}$)-Nb trilayer technology, each having a nominal equivalent square length of 4.5~$\mu$m and a critical current density, $J$, of approximately 75~A/cm$^2$. The heater resistor is placed on a region of the bottom electrode of the JJ array structure, on which an insulating oxidized niobium Nb$_{2}$O$_{5}$ surface layer is formed by anodic oxidation. An additional insulating layer of SiO$_{2}$ is deposited on top of the Nb$_{2}$O$_{5}$ surface by plasma-enhanced physical layer deposition (PECVD). This is done to provide sufficient electrical insulation between the heater and the JJs. The insertion of such an insulating layer does not deteriorate the heater efficiency, which has been validated by measurements on a set of samples with reduced SiO$_{2}$ thickness and with Nb$_{2}$O$_{5}$ insulation only. 

From the measured $I$-$V$ curve of a TCL without an applied heater current, as shown in figure~\ref{fig:fig_1} b), a maximum critical current, $I_{\rm{c}}$, of 15~$\mu$A and a normal state resistance, $R_{N}$, of 500~$\Omega$ are determined. Since the TCL is integrated into the superconducting input circuit, its normal state resistance -- when the TCL is activated -- acts as a thermal current noise source with the power spectrum density $I_{N}^{2}={}$4$k_{B}T/R_{N}$, where $k_{B}$ is the Boltzmann constant and $T$ is the temperature of the resistance $R_N$. This noise can couple to neighbouring channels via the mutual inductances of their input circuits. Therefore, $R_N$ should be sufficiently large to minimize this additional noise.

\section{Performance}

\begin{figure}[!t]
\centering
\includegraphics[width=0.91\columnwidth]{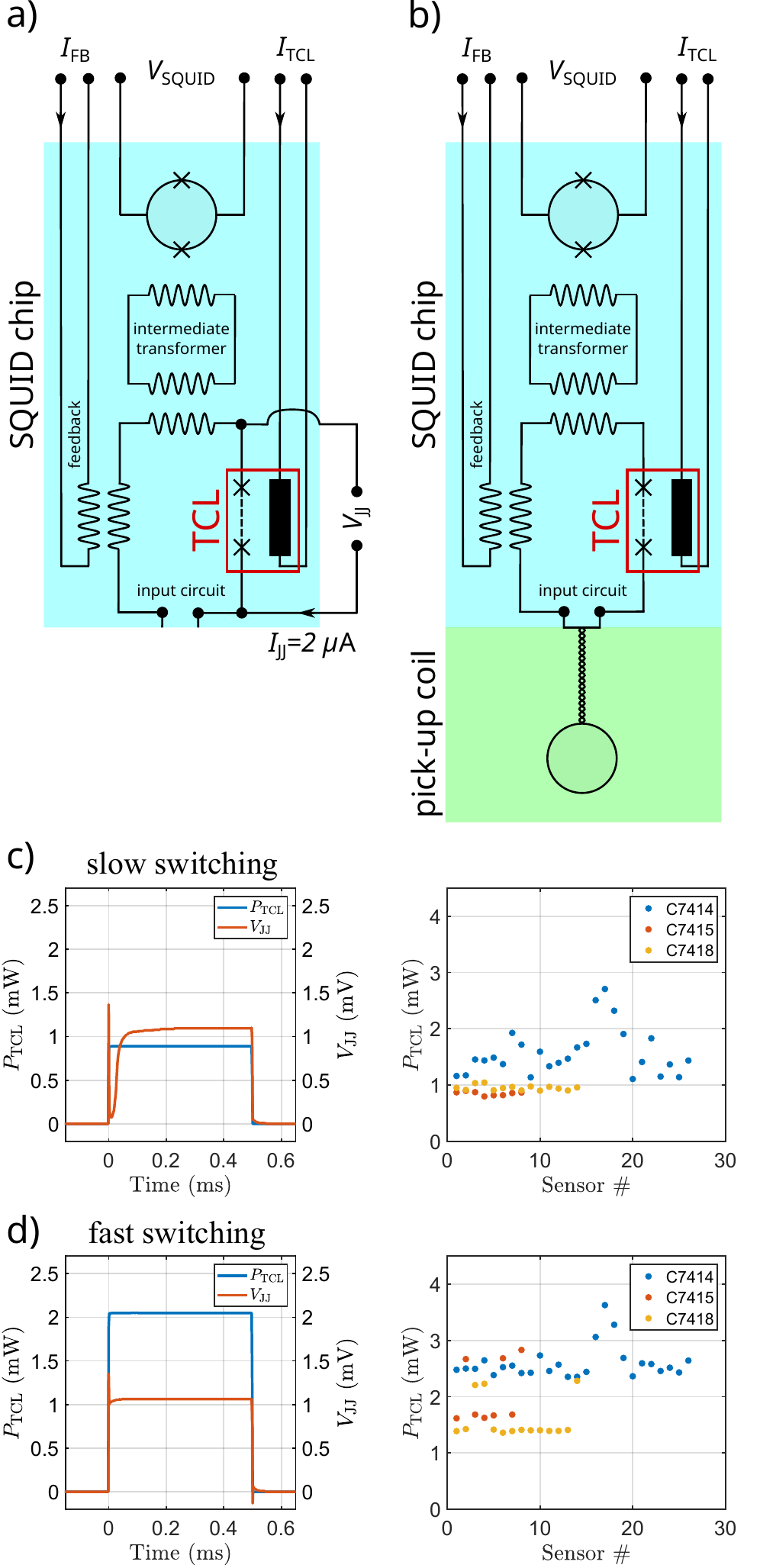}
\caption{Characterization of the TCL's switching performance. a) Observation of the voltage across the JJ array (biased with 2 $\mu$A) and b) SQUID modulation in dependence of the applied dc heater power $P_{\rm{TCL}}$. c) Slow switching and d) fast switching of the TCL using the transport measurement at 4.2~K in liquid Helium together with the statistics for different devices from separate fabrication runs. The switching time depends on the heater power and is well below 10~$\mu$s for $P_{\rm{TCL}}\gtrapprox{}$1.5~mW.}
\label{fig:fig_2}
\end{figure}

The TCL switching performance was tested at 4.2~K in liquid Helium. Two methods, as shown in figures~\ref{fig:fig_2} a) and b), were used. Firstly, the dependence of the voltage across the JJ array, $V_{\rm{JJ}}$, was measured with respect to applied dc heater power $P_{\rm{TCL}}$. Here, a small current of 2~$\mu$A was applied to the JJs in order to generate an easily measurable voltage drop across the array. In the second method, we measured the flux coupling between the input circuit and the SQUID by applying a current into the feedback coil and then observing the SQUID voltage-flux characteristics (in open loop) for different values of $P_{\rm{TCL}}$. This TCL test has the advantage of being able to be performed in the final, assembled sensor arrangement.

The results using the first approach are shown in figures~\ref{fig:fig_2} c) and d). As expected, the switching time, as defined as the time, after which all JJs are in the voltage state, depends on $P_{\rm{TCL}}$. There is some variation in the performance for different fabrication runs. In particular, the early devices (C7414) show significant spread which can be attributed to inhomogeneous fabrication parameters. For later devices (C7415 and C7418), we observed switching times of approximately 200~$\mu$s for $P_{\rm{TCL}}\approx{}$1~mW (slow switching). For $P_{\rm{TCL}}\gtrapprox{}$1.5~mW, the switching time was below 10~$\mu$s (fast switching). Note, for the coupling measurements, the switching times were shorter for the same $P_{\rm{TCL}}$ because not every JJ has to be in the normal state to prevent coupling (data not shown).

\section{Multichannel SQUID system}

\subsection{Noise performance}

\begin{figure}[!t]
\centering
\includegraphics[width=0.91\columnwidth]{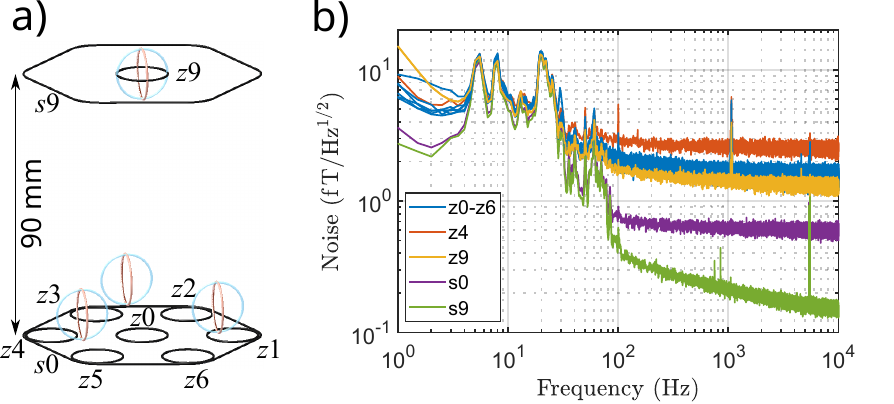}
\caption{a) Layout of the 18-channel SQUID module (to scale) featuring overlapping pick-up coils. b) Noise performance of the multichannel module. Only the $z$-channels are shown.}
\label{fig:fig_3}
\end{figure}

The current sensing SQUIDs with integrated TCL were mounted onto our 18-channel prototype, multichannel SQUID system~\cite{Storm2016} and its noise performance was investigated inside the heavily shielded room BMSR2.1 in the Physikalisch-Technische Bundesansalt (PTB) in Berlin. The system was situated centrally inside the room and operated with commercial, multichannel read-out electronics (Magnicon, MCFLL)~\cite{Bechstein2020}. The electronics were operated outside of the shielded room to avoid detrimental magnetic fields, caused by electronic components which could have worsened the field homogeneity of the heavily shielded environment. 

The noise performance of the $z$ channels is shown in figure~\ref{fig:fig_3}. As observed previously~\cite{Storm2016}, the bottom channels, $z$0-$z$6 and $s$0, are dominated by dewar noise and show a white field noise density of approximately 1.5 and 0.6~fT/Hz$^{1/2}$, respectively. The $z$4 channel has an intrinsically higher SQUID noise. The $z$9 channel noise level of 1.25~fT/Hz$^{1/2}$ stems from SQUID and electronics contributions, whereas the upturn in the $s$9 noise from 0.15~fT/Hz$^{1/2}$ at 10~kHz towards lower frequencies is generated by the $\mu$-metal walls of the BMSR2.1~\cite{Storm2017}.

\subsection{Avoiding signal distortions}

In addition to protecting the input circuit from too large currents, an important feature of our TCL is the ability to deactivate a given sensor channel by permanently 'opening' its input circuit. This would for example be necessary, if a faulty SQUID or feedback coil were to prevent feedback into the input circuit.  When the feedback works properly, any input circuit current is effectively nulled and signal cross-talk between channels from mutual input circuit couplings is minimized. If feedback into the input circuit of a given channel were not possible, while its input circuit remains fully superconducting, any field change would give rise to shielding currents which, in turn, would couple into neighbouring channels via mutual inductance (mainly due to the mutual inductance of the pick-up coils). Permanently activating the TCL of the faulty channel prevents this parasitic coupling between sensor channels. 

\begin{figure}[!t]
\centering
\includegraphics[width=0.95\columnwidth]{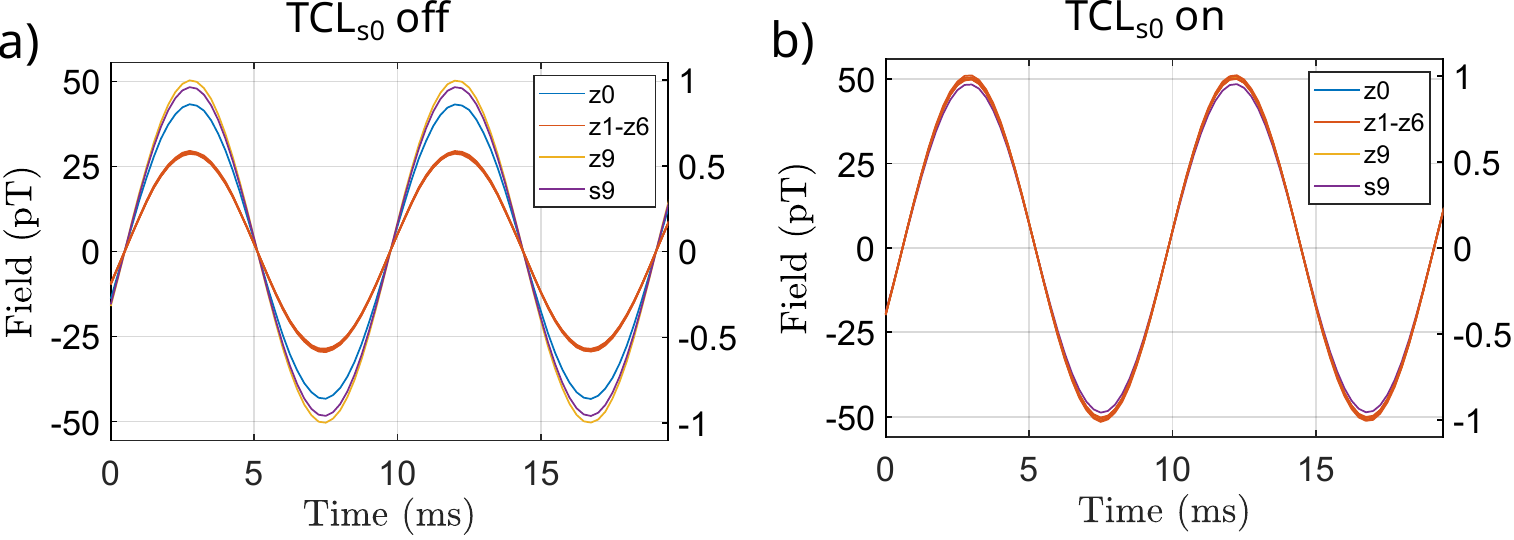}
\caption{Deactivation of the non-functioning channel $s$0. a) If the TCL of $s$0 is not activated, significant signal distortions in neighbouring channels due to shielding currents are observed. b) Activation of the TCL suppresses these shielding currents in $s$0 and the true signal size for all $z$-channels is recovered. Right axes show normalized field values.}
\label{fig:fig_4}
\end{figure}

To demonstrate this effect, a homogeneous field of $\sim$50~pT was applied in the $z$-direction, while the large bottom $s$0 pick-up coil was not operated, i.e., feedback circuit open. In figure~\ref{fig:fig_4} a), the TCL was not operated, leading to screening currents in $s$0 and causing large signal size distortions in the flux locked loop signals of the other functioning channels. A reduction of more than 40\% was observed for the strongly coupled $z$1-$z$6 coils adjacent to $s$0. For the central $z$0 coil, a smaller reduction of ~14\% was observed, whereas $s$9 was hardly affected. The distortions are removed by operating the $s$0's TCL leading to the same signal amplitude being recovered for all $z$-channels as expected for a homogeneous driving field, as shown in figure~\ref{fig:fig_4} b).

With knowledge of the pick-up coil geometry, this effect can be quantified. The flux, $\Phi_{m}$, in a pick-up coil $m$ with area $A_{m}$, due to a homogeneous field $B_{z}$, is $\Phi_{m}=B_{z}A_{m}$. When sensor $m$ is not operated, a shielding current $I_{m}=\Phi_{m}/L_{m}=B_{z}A_{m}/L_{m}$ is induced, where $L_{m}$ is the total input circuit inductance of sensor $m$. This leads to a screening flux in the pick-up coil, $n$, of $\Phi_{s,n}=M_{mn}I_{n}=M_{mn}B_{z}A_{m}/L_{m}$, where $M_{mn}$ is the mutual inductance between pick-up coils $m$ and $n$. The ratio of the screening and signal flux in $n$ is then $\Phi_{s,n}/\Phi_{n}=\left(M_{mn}/L_{m}\right)/\left(A_{m}/A_{n}\right)$. Finally, we obtain for the measured field in $n$ including screening effects due to $m$:

\begin{equation}
\centering
\label{eq:eq_1}
B_{n}^{s,m}=\left(1-\frac{\Phi_{s,n}}{\Phi_{n}}\right)B_{z}=\left(1-\frac{M_{mn}}{L_{m}}\frac{A_{m}}{A_{n}}\right)B_{z}.
\end{equation}

In table~\ref{tab:tab_1}, the calculated values $B_{n}^{s,s0}$ are shown for $z$-sensors taking screening due to $s$0 into account. The predicted values, using the computed $M_{s0\,n}$, $A_{s0}$, $A_n$ and the independently measured $L_{s0}$, are in very good agreement with observation shown in figure~\ref{fig:fig_4} a).

\begin{table}
\caption{\label{tab:tab_1}Calculated $B_{n}^{s,s0}$ with screening due to coil $s$0, $L_{s0}={}$526~nH.}
\lineup
\begin{indented}
\item[] \begin{tabular}{@{}lll}
\br
coil $n$ 	& $M_{s0\,n}$ (nH)	&	 	$B_{n}^{s,s0}$\\
\mr
	$z$0 		&				3.86 				&	 	0.86~$B_{z}$\\
$z$1-$z$6 &				11.7				&		0.575~$B_{z}$\\ 
	$s$9		& 			3.40 				& 	0.993~$B_{z}$\\ 
\br
\end{tabular}
\end{indented}
\end{table}

\subsection{Avoiding additional magnetic field gradients}

\begin{figure}[!t]
\centering
\includegraphics[width=0.95\columnwidth]{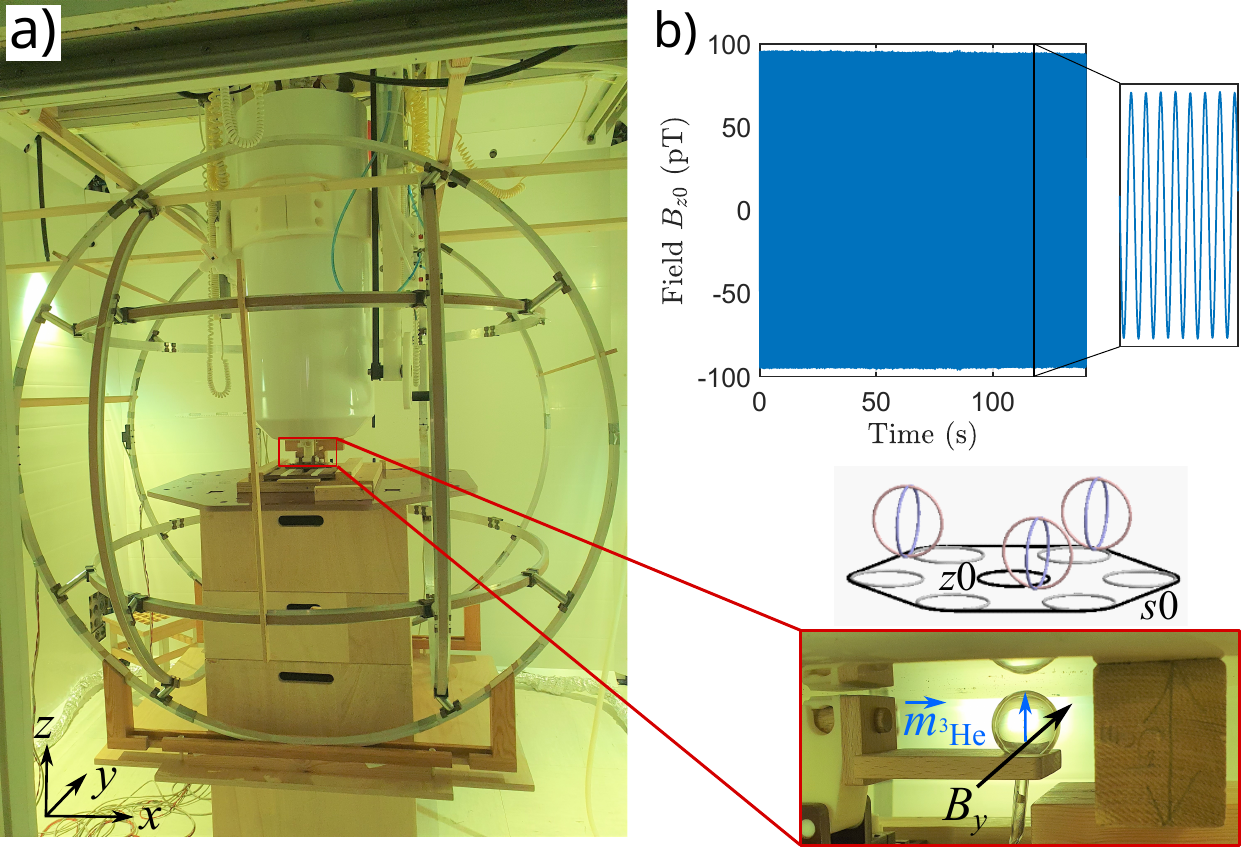}
\caption{a) Setup for high-sensitivity spin precession experiments inside the BMSR2.1. A static field of $\sim$1~$\mu$T along the $y$-direction corresponding to a Larmor frequency of 32~Hz was used. b) Detailed view showing the glass cell containing hyperpolarized $^{3}$He underneath some of the sensors. The magnetic moment $m_{^{3}\rm{He}}$ precesses around $B_{y}$ and an example of the precession signal in sensor $z0$ is given on the top.}
\label{fig:fig_5}
\end{figure}

In order to demonstrate the detrimental effects of an additional magnetic field gradient, caused by a trapped dc current in a superconducting pick-up coil, high-precision spin precession experiments of hyperpolarized $^{3}$He were carried out as shown in figure~\ref{fig:fig_5}. Noble gases, such as $^{3}$He, have intrinsic hours-long spin-lattice and spin-spin relaxation times $T_1$ and $T_2$, respectively~\cite{Gemmel2010}. However, this gaseous spin system is particularly sensitive to even minute field gradients leading to $T_2$-shortening. The sample contained 777 mbar of $^3$He at room temperature and was polarized offline by spin-exchange optical pumping~\cite{Korchak2012}. For our experimental conditions the so-called high pressure regime applies, where the spin-spin relaxation rate $1/T_2$ can be approximated to~\cite{Cates1988}:

\begin{equation}
\centering
\label{eq:eq_2}
\frac{1}{T_{2}}\approx \frac{8R^{4}\gamma^{2}\left|\nabla B_{y}\right|^{2}}{175D},
\end{equation}
where, $R={}$1.55~cm is the radius of a spherical sample cell, $\gamma=2\pi\times{}$32.43~Hz/$\mu$T is the gyromagnetic ratio of $^3$He, $\nabla B_{y}=\left(\partial B_y/\partial x \quad \partial B_y/\partial y \quad \partial B_y/\partial z\right)^{\intercal}$ is the spatial gradient of the static holding field, which in this case is in the $y$-direction, and $D={}$1.54~cm$^2$/s is the diffusion constant of the $^3$He atoms. 

\begin{figure}[!t]
\centering
\includegraphics[width=0.95\columnwidth]{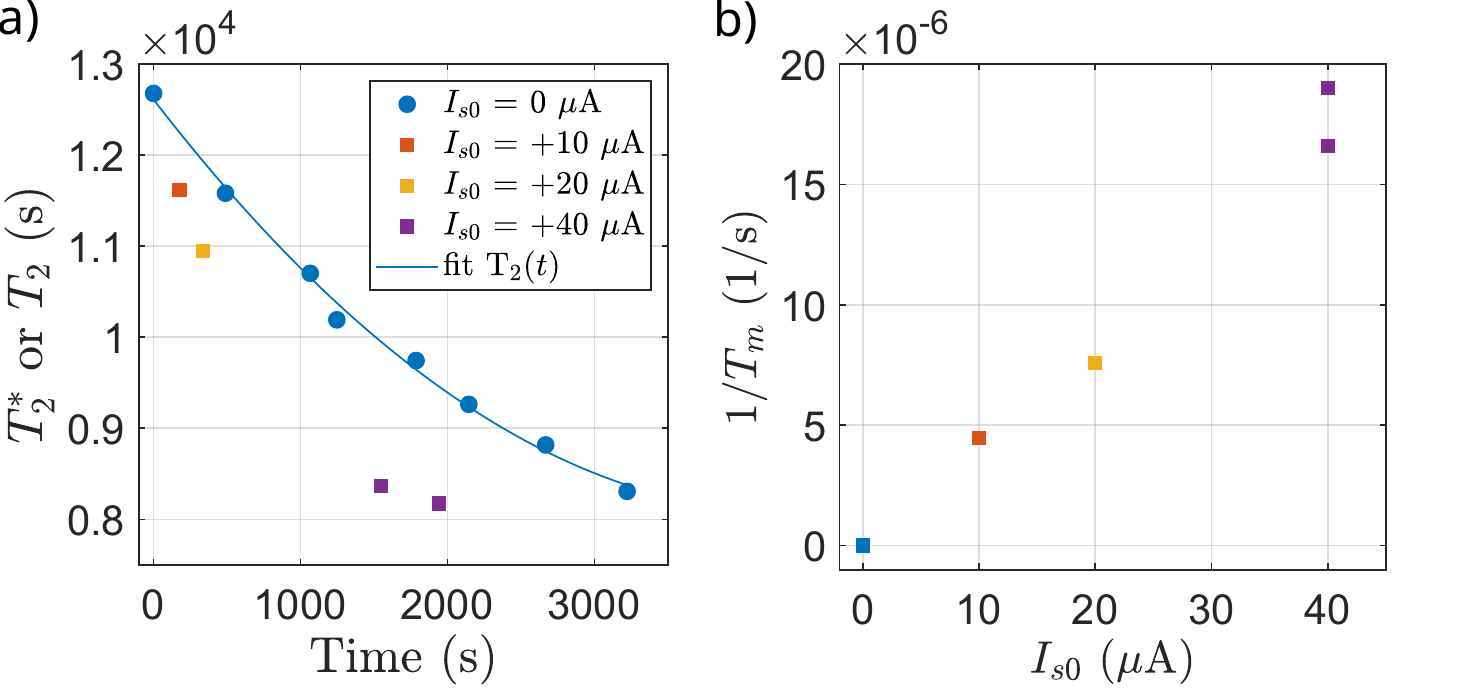}
\caption{a) Spin-spin relaxation times $T_2$ (for $I_{s0}={}$0) and $T_2^*$ (for $I_{s0}\neq{}$0) measured over a time period of about 1 hour and for different configurations. b) Contribution of the additional gradient $\nabla B_{y}$ due to a trapped current $I_{s0}$ in $s0$ quantified by $1/T_m.$}
\label{fig:fig_6}
\end{figure}

Figure~\ref{fig:fig_6} a) shows the measured $T_{2}$ for the different conditions. For $I_{s0}=0$, $T_2$ exhibits an intrinsic time dependence (blue data points), which we attribute to a time-varying gradient $\nabla B_{y}(t)$. This could be modelled phenomenologically by a second-order polynomial and likely originated from drifting background gradients or dynamic self-interactions due to the non-spherical sample. When currents of 10, 20 and 40~$\mu$A were trapped solely in the $s$0 pick-up coil, $T_2$ was markedly reduced, with the effect becoming more pronounced with increasing $I_{s0}$. The application of currents larger than the nominal critical current of 15~$\mu$A was possible due to degraded JJ characteristics. To obtain the additional $T_2$-shortening contribution due to trapped current $I_{s0}$, we introduce $T_{m}$ and express the resulting effective spin-spin relaxation rate $1/T_{2}^{*}(t)$ by:

\begin{equation}
\centering
\label{eq:eq_3}
\frac{1}{T_{2}^{*}(t)}\approx \frac{1}{T_{2}(t)}+\frac{1}{T_{m}},
\end{equation}
where the observed time dependence for $I_{s0}={}$0 is absorbed into an intrinsic $1/T_{2}(t)$ given by~(\ref{eq:eq_2}). 

Figure~\ref{fig:fig_6} b) shows a roughly linear relationship between $1/T_{m}$ and $I_{s0}$. One would expect $1/T_m\propto I_{s0}^2$, since the additional $\nabla B_y$ is linear with $I_{s0}$ and its functional dependency should follow~(\ref{eq:eq_2}). The likely cause for this discrepancy is the assumption that the time varying intrinsic gradient and the additional gradient, due to $I_{s0}$, are strictly linear in all directions and therefore, purely additive over the entire sample volume. In particular, the strict linearity of the additional gradient due to $I_{s0}$ in the close-by pick-up coil is an oversimplification. Nevertheless, significant $T_{2}$-time shortening is observed, when a dc current is trapped in $s0$  which can be prevented by actuating the TCL thereby setting $I_{s0}={}$0.

In the following, a rough estimation of the gradient $\nabla B_y$ due to the current $I_{s0}$ is presented, where we neglect the off-diagonal elements, such that $\nabla B_y \approx \partial B_y/\partial y$. Assuming a circular pick-up coil of diameter $r={}$37~mm (equivalent radius of sensor $s$0), the main gradient at the sample position underneath the coil would be along the coil's normal vector, i.e., the $z$-axis. It can be shown, $\partial B_{z,\rm{max}}/\partial z=0.75\mu_0 r^{-2}1.25^{-5/2}I_{s0}$, which occurs when $z=r/2$. From Gauss's law, div\,\textbf{B}${}={}$0, $\nabla B_y\approx \partial B_y/\partial y\approx 1/2\,\partial B_z/\partial z$, implying an expected value of $\nabla B_y/I_{s0}\approx{}$2.0~pT/(cm$\,\mu$A).
 
The resultant $1/T_m$ can then be calculated using (\ref{eq:eq_2}). For 10, 20 and 40 $\mu$A, the computed values for $1/T_m$ are 3$\times$10$^{-6}$, 12$\times$10$^{-6}$ and 48$\times$10$^{-6}$~1/s, respectively. At higher $I_{s0}$, the difference between the observed approximately linear behaviour and the predicted quadratic dependency again becomes apparent, whereas at low $I_{s0}$, there is good agreement with the measured results shown in figure~\ref{fig:fig_6} b). 

\section{Conclusion}
In conclusion, we have presented a compact, fully integrated, thermally activated current limiter, which in addition to limiting the input circuit current to about 15$~\mu$A, can be used to set the input coil critical current to zero. The TCL can be activated within $\sim$10~$\mu$s with a power of approximately 1.5~mW. In addition to pulsed fields applications, it is particularly useful for the deactivation of non-functioning channels, avoiding parasitic signal distortions due to shielding currents in their fully superconducting input circuits. In the field of high-precision spin precession experiments, detrimental gradients due to trapped dc currents can also be avoided, ensuring the best possible experimental conditions.

\ack
We would like to acknowledge fruitful discussions with Dietmar Drung.

\section*{References}
\bibliographystyle{iopart-num}

\bibliography{ASC2022_SUST_submission}

\end{document}